\documentclass[11pt,prc,aps,a4paper,groupedaddress,superscriptaddress,nofootinbib,showpacs
,preprintnumbers,onecolumn]{revtex4}
\usepackage{graphicx}
\usepackage{amsfonts}
\usepackage{amssymb}
\usepackage{natbib}
\usepackage{dcolumn}
\usepackage{bm}
\newcommand{\bwt}{\begin{widetext}}
\newcommand{\ewt}{\end{widetext}}
\newcommand{\beq}{\begin{equation}}
\newcommand{\eeq}{\end{equation}}
\newcommand{\bea}{\begin{eqnarray}}
\newcommand{\eea}{\end{eqnarray}}
\begin{document}
\title{Stellar matter with strong magnetic field within density dependent relativistic models.}
\author{A. Rabhi}
\email{rabhi@teor.fis.uc.pt}
\affiliation{Centro de F\' {\i}sica Te\'orica, Department of Physics, University of Coimbra, 3004-516 Coimbra, Portugal} 
\affiliation{Laboratoire de Physique de la Mati\`ere Condens\'ee,
Facult\'e des Sciences de Tunis, Campus Universitaire, Le Belv\'ed\`ere-1060, Tunisia}
\author{C.~Provid\^encia}
\email{cp@teor.fis.uc.pt}
\affiliation{Centro de F\' {\i}sica Te\'orica, Department of Physics, University of Coimbra, 3004-516 Coimbra, Portugal} 
\author{J.~Da~Provid\^encia}
\email{providencia@teor.fis.uc.pt}
\affiliation{Centro de F\' {\i}sica Te\'orica, Department of Physics, University of Coimbra, 3004-516 Coimbra, Portugal} 

\date{\today}
\begin{abstract}
The effect of strong magnetic fields on the equation of state  (EoS) for
compact stars described with density dependent relativistic  hadronic  models
is studied. A comparison with other mean-field relativistic models is done. It
is shown that the largest differences between models occur for low densities
and that the magnetic field affects the crust properties of star, namely its
extension. 
  
\end{abstract}
\maketitle

\section{Introduction}
The study of very asymmetric nuclear matter is presently an important issue due to the
radioactive beams which will be operating in the near future and which will
allow the investigation of a region of the nuclear matter phase space unaccessible till
recently. Asymmetric nuclear matter is of particular interest for the description of stellar matter of compact stars. 

Compact star properties depend a lot on the model used to describe the
hadronic equation of state (EoS). In particular relativistic nuclear
mean-field models~\cite{bb, glen00} are very popular to describe stellar matter because
causality will always be satisfied. The imposition of constraints, both coming
from measured star properties or from relativistic heavy ion collisions in
the laboratory, is essential to test the different models \cite{klaehn}.

Magnetars are neutron stars which may have surface magnetic fields larger that
$10^{15}$ G \cite{duncan,usov,pacz} and which were discovered at the x-ray and
$\gamma$-ray energies (for a review refer \cite{harding06}). They are identified
with the anomalous x-ray pulsars (AXP) and soft $\gamma$-ray repeaters. Taking
as reference the critical field at which the electron cyclotron energy is
equal to the electron mass $B^e_c=4.414 \times 10^{13}$ G we define $B^*=B/B^e_c$. 
It has been shown by several authors that the magnetic fields larger than
$B^*=10^5$  will affect the EoS of compact stars
\cite{chakrabarty96,broderick}. In particular field-theoretical descriptions
based on the non-linear Walecka model (NLWM) \cite{bb}  were used and several
parametrisations compared and it was shown that they had an overall similar behaviour.
Very strong magnetic fields can only occur in very  young compact stars before
the magnetic field has decayed. Recently it was shown \cite{pons08}
from a  2D calculation of the cooling of magnetized stars that the magnetics fields and
Joule heating have an important effect of maintaining compact stars warm for a
longer time. This kind of simulations need the EoS of the crust. It is,
therefore, important to make a study that shows when should the magnetic field
be taken into account explicitly in the EoS of the crust. An unstable
region in a wider density range  will correspond to a larger crust and the
properties of the star depending on the crust will be affected. It should,
however, be pointed out the estimated surface magnetic fields of detected  magnetars  is
obtained  assuming that the lost of angular momentum is entirely due to dipolar
radiation of magnetars, and until recently, the strongest estimated magnetic
field is of the order of $B^*=10^2$ and was detected in  a quite young
star, SGR 1806-20 \cite{sgr}.

At low densities relativistic models with constant coupling parameters
have different behaviour from density dependent relativistic hadron models (DDRH)~\cite{fuchs,tw}. 
These models have density dependent coupling parameters  and
have originally been parametrised so that relativistic Dirac Brueckner
Hartree-Fock (DBHF) calculations for nuclear matter were reproduced~\cite{dbhf}.
 Within DDRH models the symmetry energy does not increase 
linearly with density as parametrizations of the NLWM, like the 
ones studied in~\cite{broderick}, and show a behavior closer to non-relativistic models,
either some of the recent parametrizations of the   Skyrme interaction like
SLy230a~\cite{chabanat} or NRAPR~\cite{steiner05} or variational microscopic
calculations~\cite{apr}.
Since nuclear matter is composed of two different fluids, namely protons and 
neutrons, the liquid gas phase transition can lead to an isospin distillation 
phenomenon which has been confirmed experimentally~\cite{xu00}.
In~\cite{cp07,magno07} it was shown that the distillation effect was described both by NLWM and  DDRH
models, but DDRH models did not predict an effect so strong as the first ones. 
Moreover, it was shown that DDRH models have at subsaturation densities a behaviour similar to non-relativistic
models, namely models with Skyrme forces~\cite{abmp06,del07}. 
We would like to test these models under different conditions namely
$\beta$-equilibrium matter under strong magnetic fields. Stellar matter, as
found in compact stars, under strong magnetic fields has already been studied
before by many authors~\cite{chakrabarty96,yuan,broderick,shen06}.  

The authors of Ref.~\cite{kubis97,liu} have stressed the importance
of including the scalar isovector virtual $\delta (a_0(980))$ field in 
hadronic effective field theories when asymmetric nuclear matter is studied.  Its presence 
introduces in the isovector channel the structure of relativistic
interactions, where a balance between a scalar (attractive) and a vector
(repulsive) potential exists. The introduction of the $\delta$ meson mainly
affects the behaviour of the system at high densities, when, due to
Lorentz contraction, its contribution is reduced, leading to a harder
EoS at densities larger than $\sim 1.5\, \rho_0$~\cite{liu}. 
In~\cite{mp04} the effect of this meson on the properties of
compact stars were studied and it was shown that the EoS of hadronic matter
would become stiffer with its presence. In relativistic models there is a proton-neutron mass splitting only
if the scalar isovector $\delta$-meson is included. This occurs for
the DDRH$\delta$  parametrization we consider. 
For this model  $M^*_n<M_p^*$ in neutron rich nuclear matter~\cite{brito07}. 
A similar behavior is predicted by the Skyrme interaction SLy230a  but an opposite 
behavior is obtained with other parametrizations of the Skyrme interaction~\cite{baran05}.
In a recent work~\cite{baoli07} the behavior of the proton-neutron mass 
splitting in different relativistic nuclear models was analysed and it was 
shown that some point-coupling models without mesons~\cite{fkvw} 
predict larger neutron masses in neutron rich matter. The proton-neutron 
mass splitting is  a present topic of discussion and the forecoming experiments 
with  radioactive beams will allow the clarification of this point. We will investigate 
whether this mass spitting has some effect on stellar matter with strong magnetic fields.

In the present paper we will study the behaviour of stellar matter described within DDRH, 
both with~\cite{gaitanos} and without~\cite{tw} the isovector-scalar $\delta$-meson, 
under very strong magnetic fields. The results are compared with previously studied models, 
the parametrisations GM3~\cite{gm91} and TM1~\cite{tm1} of NLWM. These two models 
including strong magnetic fields, have been discussed in~\cite{broderick} and~\cite{shen06}.

In section II we make a brief review of the model and EoS under the effect of
a magnetic field. In section III the formalism is generalised to include the
$\delta$-meson. Results are discussed in section IV and conclusions are drawn
in section V. 

\section{The formalism}
For the description of the EoS of neutron star matter, we employ a field-theoretical approach 
in which the baryons (neutrons,  n, and protons, p) interact via the exchange of $\sigma-\omega-\rho$ 
mesons in the presence of a uniform magnetic field $B$ along the $z$-axis. The Lagrangian density 
of the relativistic TW model~\cite{fuchs,tw} can be written as  
\beq
{\cal L}= \sum_{b=n, p}{\cal L}_{b} + {\cal L}_{m}+ \sum_{l=e, \mu}{\cal L}_{l}
\label{lan}
\eeq
The baryons ($b$=$n$, $p$), leptons ($l$=$e$, $\mu$), and mesons ($\sigma$, $\omega$ and  $\rho$) Lagrangians are given 
by
\bwt
\bea
{\cal L}_{b}&=&\bar{\Psi}_{b}\left(i\gamma_{\mu}\partial^{\mu}-q_{b}\gamma_{\mu}A^{\mu}- 
m_{b}+\Gamma_{\sigma}\sigma
-\Gamma_{\omega}\gamma_{\mu}\omega^{\mu}-\frac{1}{2}\Gamma_{\rho}\tau_{3 b}\gamma_{\mu}\rho^{\mu}
-\frac{1}{2}\mu_{N}\kappa_{b}\sigma_{\mu \nu} F^{\mu \nu}\right )\Psi_{b} \cr
{\cal L}_{l}&=& \bar{\psi}_{l}\left(i\gamma_{\mu}\partial^{\mu}-q_{l}\gamma_{\mu}A^{\mu}
-m_{l}\right )\psi_{l}  \cr
{\cal L}_{m}&=&\frac{1}{2}\partial_{\mu}\sigma \partial^{\mu}\sigma
-\frac{1}{2}m^{2}_{\sigma}\sigma^{2}
+\frac{1}{2}m^{2}_{\omega}\omega_{\mu}\omega^{\mu}
-\frac{1}{4}\Omega^{\mu \nu} \Omega_{\mu \nu}  \cr
&-&\frac{1}{4} F^{\mu \nu}F_{\mu \nu}
+\frac{1}{2}m^{2}_{\rho}\rho_{\mu}\rho^{\mu}-\frac{1}{4}  P^{\mu \nu}P_{\mu \nu}
\label{lagran}
\eea
\ewt
where $\Psi_{b}$ and $\psi_{l} $ are the baryon and lepton Dirac fields, respectively. The nucleon mass and isospin projection for the proton and neutrons
are denoted by $m_{b}$ and $\tau_{3 b}=\pm 1$, respectively. The mesonic and electromagnetic field strength tensors are given by 
their usual expressions: $\Omega_{\mu \nu}=\partial_{\mu}\omega_{\nu}-\partial_{\nu}\omega_{\mu}$, $P_{\mu 
\nu}=\partial_{\mu}\rho_{\nu}-\partial_{\nu}\rho_{\mu}$, and  $F_{\mu \nu}=\partial_{\mu}A_{\nu}-\partial_{\nu}A_{\mu}$. The nucleon anomalous magnetic moments are introduced via the coupling of the baryons to the electromagnetic field tensor with $\sigma_{\mu \nu}=\frac{i}{2}\left[\gamma_{\mu},  \gamma_{\nu}\right] $ and strength $\kappa_{b}$ with $\kappa_{n}=-1.91315$ for the neutron and $\kappa_{p}=1.79285$ for the proton, respectively. The electromagnetic field is assumed to be externally generated (and thus has no associated field equation), and only frozen-field configurations will be considered. The density dependent strong interaction couplings are denoted by $\Gamma$, the electromagnetic couplings by $q$ and the nucleon, mesons and leptons masses by $m$. The parameters of the model are the nucleon mass $m_b=939$ MeV, the masses of mesons $m_\sigma$, $m_\omega$, $m_\rho$ and the density dependent coupling parameters which are adjusted in order to reproduce some of the nuclear matter bulk properties and relations with DBHF calculations~\cite{dbhf}, using the following parametrisation
\beq
\Gamma_{i}(\rho)=\Gamma_{i}(\rho_{sat})f_{i}(x),\quad i=\sigma, \omega
\label{gam1}
\eeq
with
\beq
f_{i}(x)=a_{i}\frac{1+b_{i}\left(x+d_{i}\right)^{2}}{1+c_{i}\left(x+d_{i}\right)^{2}}
\label{gam2}
\eeq
where $x={\rho}/{\rho_{sat}}$ and
\beq
\Gamma_{\rho}(\rho)=\Gamma_{\rho}(\rho_{sat})\exp\left[ -a_{\rho}(x-1)\right], 
\eeq
with the values of the parameters $m_i$, $\Gamma_{i}$, $a_{i}$, $b_{i}$, $c_{i}$ and $d_{i}$, $i=\sigma, \omega, \rho$ 
given in \cite{tw}. Other possibilities for these parameters are also found in the literature~\cite{hbt}. 

The field equations of motion follow from the Euler-Lagrange equations. From the Lagrangian density in Eq.~(\ref{lan}), we 
obtain the following meson field equations in the mean-field approximation 
\bea
m^{2}_{\sigma}\left\langle \sigma \right\rangle &=&\Gamma_{\sigma}\left(\rho^{s}_{p}+\rho^{s}_{n}\right) 
=\Gamma_{\sigma}\rho^{s}\label{mes1} \\
m^{2}_{\omega} \left\langle \omega^{0}\right\rangle  &=& \Gamma_{\omega}\left(\rho^v_{p}+\rho^v_{n}\right)= 
\Gamma_{\omega}\rho_{b}\label{mes2} \\
m^{2}_{\rho} \left\langle \rho^{0}\right\rangle  &=& \frac{1}{2}\Gamma_{\rho}\left(\rho^v_{p}-\rho^v_{n}\right) =\frac{1}
{2}\Gamma_{\rho}\rho_{3}\label{mes3}
\eea
and the Dirac equations for nucleons and leptons are given by 
\bea
(i\gamma_{\mu}\partial^{\mu}-q_{b}\gamma_{\mu}A^{\mu}-(m_{b}
-\Gamma_{\sigma}\sigma)-\Gamma_{\omega}\gamma_{0}\omega^{0} \cr
-\frac{1}{2}\Gamma_{\rho}\tau_{3 b}\gamma_{0}\rho^{0}-\gamma_{0}\Sigma^{R}_{0} 
-\frac{1}{2}\mu_{N}\kappa_{b}\sigma_{\mu \nu} F^{\mu \nu}) \Psi_{b}&=&0 \label{MFbary}\\
\left(i\gamma_{\mu}\partial^{\mu}-q_{l}\gamma_{\mu}A^{\mu}-m_{l} \right) \psi_{l}&=&0 \label{MFlep}
\eea
where the effective baryon masses are given by 
\beq
m^{*}_{b}=m_{b}-\Gamma_{\sigma}\sigma \label{effmass}
\eeq 
and $\rho^{s}$ is the scalar number density.  In charge-neutral, $\beta$-equilibrated  matter, the conditions 
\beq
\mu_{n}-\mu_{p}=\mu_{e}=\mu_{\mu},
\label{beta}
\eeq
and
\beq
\rho^{v}_{p}=\rho^{v}_{e}+\rho^{v}_{\mu} \label{neutra}
\label{dens}
\eeq
should be satisfied.
  
The energy spectra for protons, neutrons and leptons (electrons and muons) are given by
\bea
E^{p}_{\nu, s}&=& \sqrt{k^{2}_{z}+\left(\sqrt{m^{* 2}_{p}+2\nu q_{p}B}-s\mu_{N}\kappa_{p}B \right) 
^{2}}+\Gamma_{\omega} \omega^{0}+\frac{1}{2}\Gamma_{\rho}\rho^{0}+\Sigma^{R}_{0} \label{enspc1}\\
E^{n}_{s}&=& \sqrt{k^{2}_{z}+\left(\sqrt{m^{* 2}_{n}+k^{2}_{x}+k^{2}_{y}}-s\mu_{N}\kappa_{n}B 
\right)^{2}}+\Gamma_{\omega} \omega^{0}-\frac{1}{2}\Gamma_{\rho}\rho^{0}+\Sigma^{R}_{0}\label{enspc2} \\
E^{l}_{\nu, s}&=& \sqrt{k^{2}_{z}+m_{l}^{2}+2\nu |q_{l}| B}\label{enspc3}
\eea
where $\nu=n+\frac{1}{2}-sign(q)\frac{s}{2}=0, 1, 2, \ldots$ enumerates the Landau levels of the fermions with electric 
charge $q$, the quantum number $s$ is $+1$ for spin up and $-1$ for spin down
cases, and the rearrangement term is given by 
\beq
\Sigma^{R}_{0}=\frac{\partial \Gamma_{\omega}}{\partial \rho}\rho_b\omega_{0}+\frac{\partial \Gamma_{\rho}}{\partial 
\rho}\rho_{3}\frac{\rho_{0}}{2}-\frac{\partial \Gamma_{\sigma}}{\partial \rho}\rho^s\sigma.
\eeq
 
The expressions of the scalar and vector densities for protons and neutrons are given by~\cite{broderick}
\bea
\rho^{s}_{p}&=&\frac{q_{p}Bm^{*}_{p}}{2\pi^{2}}\sum_{\nu=0}^{\nu_{\mbox{\small max}}}\sum_{s}\frac{\sqrt{m^{* 2}_{p}+2\nu 
q_{p}B}-s\mu_{N}\kappa_{p}B}{\sqrt{m^{* 2}_{p}+2\nu q_{p}B}}\ln\left|\frac{k^{p}_{F,\nu,s}+E^{p}_{F}}
{\sqrt{m^{* 2}_{p}+2\nu q_{p}B}-s\mu_{N}\kappa_{p}B} \right|, \cr
\rho^{s}_{n}&=&\frac{m^{*}_{n}}{4\pi^{2}}\sum_{s} \left[E^ {n}_{F}k^{n}_{F, s}-\bar{m}^{2}_{n}\ln\left|
\frac{k^{n}_{F,s}+E^{n}_{F}}{\bar{m}_{n}} \right|\right],  \cr
\rho^{v}_{p}&=&\frac{q_{p}B}{2\pi^{2}}\sum_{\nu=0}^{\nu_{\mbox{\small max}}}\sum_{s}k^{p}_{F,\nu,s},  \cr
\rho^{v}_{n}&=&\frac{1}{2\pi^{2}}\sum_{s}\left[ \frac{1}{3}\left(k^{n}_{F, s}\right) ^{3}-\frac{1}
{2}s\mu_{N}\kappa_{n}B\left(\bar{m}_{n}k^{n}_{F,s}+E^{n 2}_{F}\left(\arcsin\left( \frac{\bar{m}_{n}}
{E^{n}_{F}}\right) -\frac{\pi}{2} \right)  \right) \right] 
\eea
and the vector densities for leptons are given by
\beq
\rho^{v}_{l}=\frac{|q_{l}|B}{2\pi^{2}}\sum_{\nu=0}^{\nu_{\mbox{\small max}}}\sum_{s}k^{l}_{F,\nu,s}
\eeq
where $k^{p}_{F, \nu, s}$, $ k^{n}_{F, s}$ and $k^{l}_{F, \nu, s}$ are the Fermi momenta of protons, neutrons and 
leptons, which are related to the Fermi energies $E^{p}_{F}$, $E^{n}_{F}$ and  $E^{l}_{F}$ as
\bea
k^{p 2}_{F,\nu,s}&=&E^{p 2}_{F}-\left[\sqrt{m^{* 2}_{p}+2\nu q_{p}B}-s\mu_{N}\kappa_{p}B\right] ^{2} \cr
k^{n 2}_{F,s}&=&E^{n 2}_{F}-\bar{m}^{2}_{n} \cr
k^{l 2}_{F,\nu,s}&=&E^{l 2}_{F}-\left(m^{2}_{l}+2\nu |q_{l}| B\right) , \quad l=e, \mu 
\eea
with
\beq
\bar{m}_{n}=m^{*}_{n}-s\mu_{N}\kappa_{n}B.\label{barm}
\eeq
The summation in $\nu$ in the above expressions terminates at $\nu_{max}$,
 the largest value of $\nu$ for which the square of Fermi momenta of
 the particle is still positive and which corresponds to the closest
 integer from below defined by the ratio
$$\nu_{max}=\left[\frac{(E^i_F)^2-m_i^2}{2 |q_i|\, B}\right],\quad \mbox{leptons}$$
$$\nu_{max}=\left[\frac{(E^p_F+s\,\mu_N\,\kappa_p\,B)^2-{m_p^*}^2}{2 |q_p|\,
    B}\right], \quad \mbox{protons}.$$
The chemical potentials of baryons and leptons are defined as 
\bea
\mu_{b}&=& E^{b}_{F}+\Gamma_{\omega}\omega^{0}+\frac{1}{2}\Gamma_{\rho}\tau_{3 b}\rho^{0}+\Sigma^{R}_{0} \\
\mu_{l} &=& E^{l}_{F}=\sqrt{k^{l 2}_{F,\nu,s}+m^{2}_{l}+2\nu |q_{l}| B}.
\eea
We solve the coupled Eqs.~(\ref{mes1})-(\ref{neutra}) self-consistently at a given baryon density in the presence of strong magnetic fields. The energy 
density of neutron star matter is given by (the index "m" refers to matter) 
\beq
\varepsilon_{m}=\sum_{b=p,n} \varepsilon_{b}+\sum_{l=e,\mu}\varepsilon_{l}+\frac{1}
{2}m^{2}_{\sigma}\sigma^{2}+\frac{1}{2}m^{2}_{\omega}\omega^{2}_{0}+\frac{1}{2}m^{2}_{\rho}\rho^{2}_{0}
\eeq 
where the energy densities of nucleons and leptons have the following forms
\bea
\varepsilon_{p}&=&\frac{q_{p}B}{4\pi^ {2}}\sum_{\nu=0}^{\nu_{\mbox{\small max}}}\sum_{s}\left[k^{p}_{F,\nu,s}E^{p}_{F}
+\left(\sqrt{m^{* 2}_{p}+2\nu q_{p}B}-s\mu_{N}\kappa_{p}B\right) ^{2} 
\ln\left|\frac{k^{p}_{F,\nu,s}+E^{p}_{F}}{\sqrt{m^{* 2}_{p}+2\nu q_{p}B}-s\mu_{N}\kappa_{p}B} \right|\right] , \cr
\varepsilon_{n}&=&\frac{1}{4\pi^ {2}}\sum_{s}\bigg[\frac{1}{2}k^{n}_{F, s}E^{n 3}_{F}-\frac{2}
{3}s\mu_{N}\kappa_{n} B E^{n 3}_{F}\left(\arcsin\left(\frac{\bar{m}_{n}}{E^{n}_{F}} \right)-\frac{\pi}
{2}\right)-\left(\frac{1}{3}s\mu_{N}\kappa_{n} B +\frac{1}{4}\bar{m}_{n}\right) \cr
&&\left(\bar{m}_{n}k^{n}_{F, s}E^{n}_{F}+\bar{m}^{3}_{n}\ln\left|\frac{k^{n}_{F,s}+E^{n}_{F}}{\bar{m}_{n}} 
\right|\right) \bigg] \cr
\varepsilon_{l}&=&\frac{|q_{l}|B}{4\pi^ {2}}\sum_{\nu=0}^{\nu_{\mbox{\small max}}}\sum_{s}\left[k^{l}_{F,\nu,s}E^{l}_{F}
+\left(m^{2}_{l}+2\nu |q_{l}|B\right) 
\ln\left|\frac{k^{l}_{F,\nu,s}+E^{l}_{F}}{\sqrt{m^{2}_{l}+2\nu |q_{l}| B}} \right|\right] 
\eea
The pressure of the system is obtained from the expression
\beq
P_{m}=\sum_{i}\mu_{i}\rho^{i}_{v}-\varepsilon_{m}=\mu_{n}\rho_{b}-\varepsilon_{m}
\label{press}
\eeq
where the charge neutrality and $\beta$-equilibrium conditions are used to get the last equality. Note that the contribution from 
electromagnetic fields to the energy density and pressure,  $\displaystyle\varepsilon_{f}=P_{f}=\frac{B^{2}}{8 \pi}$, should 
be taken into account in the calculation of the EoS. 

\section{Including isovector-scalar mesons}
To investigate the influence of the $\delta$-meson we have included in the TW
model the isovector-scalar meson term~\cite{gaitanos}, which 
has density dependent coupling parameters. The Lagrangian density reads
\beq
{\cal L}= \sum_{b=n, p}{\cal L'}_{b}+ {\cal L'}_{m}+ \sum_{l=e, \mu}{\cal L}_{l},
\label{lan2}
\eeq
where the baryon ($b$=$n$, $p$), lepton ($l$=$e$, $\mu$), and meson ($\sigma, \omega, \rho\: \hbox{and}\:\delta$) Lagrangian 
are given by
\bea
{\cal L'}_{b}&=&{\cal L}_{b}+\bar{\Psi}_{b}\Gamma_{\delta}\vec{\tau}_{b}\cdot\vec{\delta}\Psi_{b} \cr
{\cal L'}_{m}&=&{\cal L}_{m}+{\cal L}_{\delta}  \cr
{\cal L}_{\delta}&=&\frac{1}{2}\left( \partial_{\mu}\vec{\delta} \,\partial^{\mu}\vec{\delta}
-m^{2}_{\delta}\vec{\delta}^{2}\right),
\eea
with ${\cal L}_{b}$ and ${\cal L}_{m}$  defined in Eq~(\ref{lagran}).  $\Gamma_{\delta}$ and $m_{\delta}$  
are, respectively, the coupling constant of the $\delta$ meson with the nucleons and its mass. For  $\Gamma_{\sigma}$ and 
$\Gamma_{\omega}$ we take the parametrisations given in Eqs.~(\ref{gam1}) and (\ref{gam2}). For $\Gamma_{\rho}$ and 
$\Gamma_{\delta}$, we use the parametrisation~\cite{abmp04}
\beq
\Gamma_{i}(\rho)=\Gamma_{i}(\rho_{sat})f_{i}(x), \quad x=\frac{\rho}{\rho_{sat}}
\label{gam3}
\eeq
with
\beq
f_{i}(x)=a_{i}\exp\left[ -b_{i}(x-1)\right]-c_{i}\left(x-d_{i}\right), \quad i=\rho, \delta
\label{gam4}
\eeq
and the parameters $a_{i}$, $b_{i}$, $c_{i}$ and $d_{i}$ defined in Table~\ref{table1}.
\begin{table}[htb]
\begin{tabular}{cccccc}
\hline
\hline
i & $\Gamma_{i}$ & $a_{i}$ & $b_{i}$ & $c_{i}$ & $d_{i}$  \\
\hline
$\rho$ & 11.727  & 0.095268  & 2.171 & 0.05336 & 17.8431 \\
\hline
$\delta$ & 7.58963 & 0.01984 & 3.4732 & -0.0908 &-9.811 \\
\hline
\hline
\end{tabular}
\caption{Parameters of the DDRH$\delta$ model.}
\label{table1}
\end{table}   

From the Lagrangian density in Eq.~(\ref{lan2}), we obtain the meson field 
equations~(\ref{mes1}),~(\ref{mes2}),~(\ref{mes3}) plus an equation for the $\delta$-meson in the mean-field 
approximation 
\beq
m^{2}_{\delta}\left\langle \delta_{3}\right\rangle = \Gamma_{\delta}\left(\rho^{s}_{p}-\rho^{s}_{n}\right) 
=\Gamma_{\delta}\rho^{s}_{3}
\eeq
and the Dirac equations for nucleons  are given by 
\bea
\bigg(i\gamma_{\mu}\partial^{\mu}-q_{b}\gamma_{\mu}A^{\mu}-(m_{b}
-\Gamma_{\sigma}\sigma-\Gamma_{\delta}\tau_{3 b}\delta_{3})
-\Gamma_{\omega}\gamma_{0}\omega^{0} \cr
-\frac{1}{2}\Gamma_{\rho}\tau_{3 b}\gamma_{0}\rho^{0}-\gamma_{0}\Sigma^{R}_{0} 
-\frac{1}{2}\mu_{N}\kappa_{b}\sigma_{\mu \nu} F^{\mu \nu}\bigg) \Psi_{b}&=&0. 
\eea
The effective baryon masses, in this case, are given by 
\beq
m^{*}_{b}=m_{b}-\Gamma_{\sigma}\sigma- \tau_{3 b}\Gamma_{\delta}\delta_{3},
\eeq 
and differ for protons and neutrons.
In charge-neutral, $\beta$-equilibrated  matter, the conditions  Eq.~(\ref{beta}) and Eq.~(\ref{dens}) apply.
  
The energy spectra for protons, neutrons and leptons are given by 
Eqs.~(\ref{enspc1}),~(\ref{enspc2}) and~(\ref{enspc3}) with the rearrangement term, in this case,  given by
\beq
\Sigma^{R}_{0}=\frac{\partial \Gamma_{\omega}}{\partial \rho}\rho_b\omega_{0}+\frac{\partial \Gamma_{\rho}}{\partial 
\rho}\rho_{3}\frac{\rho_{0}}{2}-\frac{\partial \Gamma_{\sigma}}{\partial \rho}\rho^s\sigma
+\frac{\partial \Gamma_{\delta}}{\partial \delta}\rho^s_3\delta_{3}.
\eeq

The expressions of the scalar and vector densities for protons and neutrons, Fermi momenta, chemical potentials still hold. The energy density of neutron star matter is, now,  given by
\beq
\varepsilon_{m}=\sum_{b=p,n} \varepsilon_{b}+\sum_{l=e,\mu}\varepsilon_{l}+\frac{1}
{2}m^{2}_{\sigma}\sigma^{2}+\frac{1}{2}m^{2}_{\omega}\omega^{2}_{0}+\frac{1}
{2}m^{2}_{\rho}\rho^{2}_{0}+\frac{1}{2}m^{2}_{\delta}\delta^{2}_{3}
\eeq 
including an extra term for the $\delta$-meson. For the pressure Eq.~(\ref{press}) holds.

\section{Results and discussion}

In the present section we  discuss the  EoS of stellar matter obtained within TW and DDRH$\delta$
and compare them with other previously studied models GM3 and TM1. We will pay a
special attention to the behaviour of the EoS at subsaturation densities in order to
understand how could strong magnetic fields affect the crust of compact stars
by extending or reducing the non-homogeneous phase.
\begin{figure}[b]
\centering
\includegraphics[width=0.75\linewidth,angle=-90]{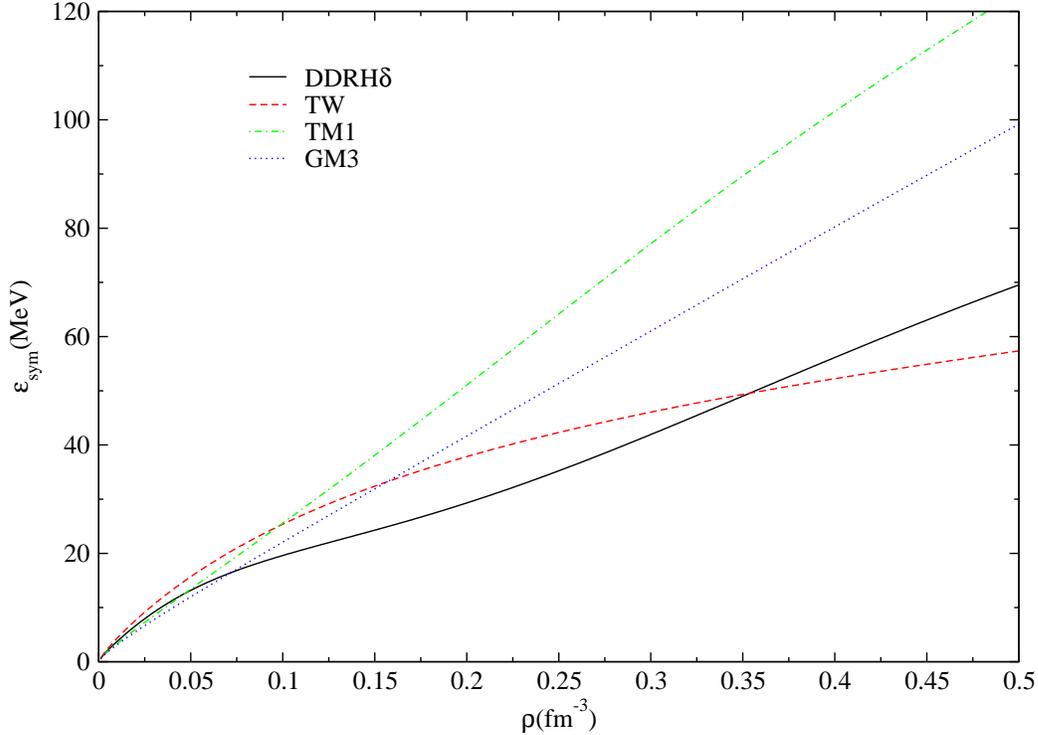}
\caption{(Color online) The symmetry energy of the models  under study.}
\label{esym}
\end{figure}

The properties of the isovector channel of the models have an important
role on the properties of the  EoS of very asymmetric matter and therefore we
plot, for reference, in Fig.~\ref{esym} the symmetry energy of the four models
under study. It is seen that the models have quite different behaviours at
large densities and this will reflect itself on the EoS. In fact the largest
proton fractions occur for the models with the largest symmetry energy since a
very asymmetric system will be energetically defavoured. While the DDRH$\delta$
is the one with the lower symmetry energy at lower densities, around two times the saturation density its symmetry energy crosses the corresponding curve for the TW. This is due to the saturation of the $\delta$ field as discussed in~\cite{liu}. We may expect that this effect will have influence on the properties of the EoS. GM3 and TM1 have quite high symmetry energy $\epsilon_{\mbox{\small sym}}$ which originates a large proton fraction at subsaturation densities and allows for the direct URCA process. We will see, however that the proton fraction is determined also by the magnitude of the effective mass.

\begin{figure}[ht]
\centering
\includegraphics[width=0.75\linewidth,angle=-90]{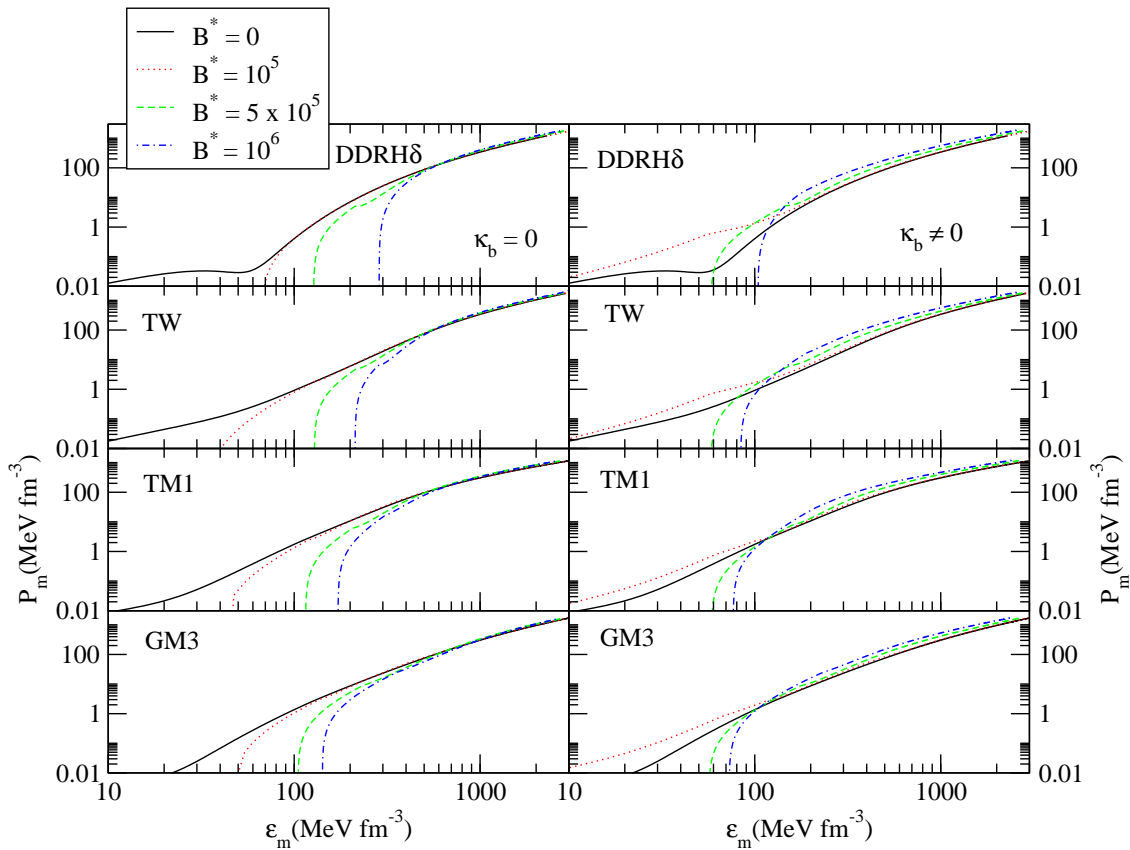}
\caption{(Color online)  EoS for stellar matter without (left) and with (right) 
the nucleon anomalous magnetic moment. Several models are considered.}
\label{eos}
\end{figure}

In Fig.~\ref{eos} we compare the EoS of TM1, GM3, TW and DDRH$\delta$ for
several magnetic field intensities. As discussed in \cite{broderick} the
magnetic field makes the EoS softer when the anomalous magnetic moment (AMM) of the nucleons
is not considered. For $B^*=10^5$ the strongest effect occurs at low densities
typical of the crust of the star. It is therefore important to investigate the
influence of the  magnetic field on the crust properties. For $B=0$ it is
shown that DDRH$\delta$ presents a small liquid-gas phase transition. This is
certainly due to the small symmetry energy this model has at low
densities. From a qualitative point of view all models behave similarly, also
when the AMM is included. In this case the EoS becomes stiffer as the magnetic field increases except for the small densities. Inparticular the curve for $B^*=10^6 $ gets softer than the $B=0$ EoS for
energy densities smaller than 70 MeV/fm$^3$, when the effect of the Landau
quantisation is stronger than the AMM contribution.
\begin{figure}[htb]
\centering
\includegraphics[width=0.75\linewidth,angle=-90]{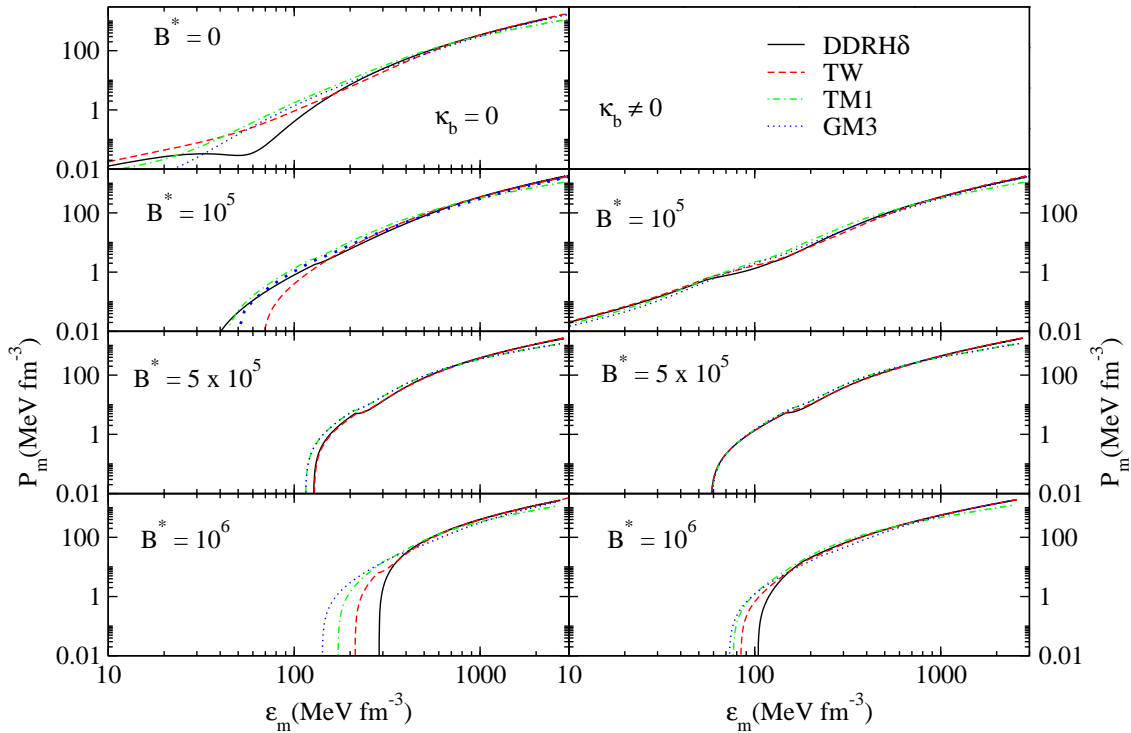}
\caption{(Color online)  EoS for stellar matter without (left) and with (right) the nucleon anomalous magnetic moment. Several magnetic fields are considered.}
\label{eos2}
\end{figure}

In order to have a more quantitative comparison, in  Fig. \ref{eos2} we plot for each magnetic field the four models. If the anomaly is not considered the largest differences occur for the lower densities: at $B=0$ TM1 and GM3 are the
softest at low densities and become the hardest at high densities; DDRH$\delta$ is the softest EoS at intermediate energies. The $B=0$ behaviour is determined by the symmetry energy, and the trend of the EoS follows the relative behaviour of the symmetry energy curves. However the relative stiffness of the EoS depends on the intensity of $B$.  Contrary to the  lower values of $B^*$,\,0 and $10^5$, for the highest value GM3 is stiffer than TM1 and for $B^*=5\times 10^5$ both EoS have very similar behaviours. Although DDRH$\delta$ EoS is  softer than the TW EoS below  $\epsilon=200$ MeV/fm$^3$ for small and large magnetic fields, for $B^{*}=$ $10^5$ and $5\times 10^5$ it becomes harder or similar. If the anomaly is included the differences between the models are much smaller mainly for  $B^*< 10^6$. For these values, models coincide at intermediate and high densities. For the low densities, differences arise  for $B^*\sim 10^6$ or larger. 
\begin{figure}[htb]
\centering
\includegraphics[width=0.75\linewidth,angle=-90]{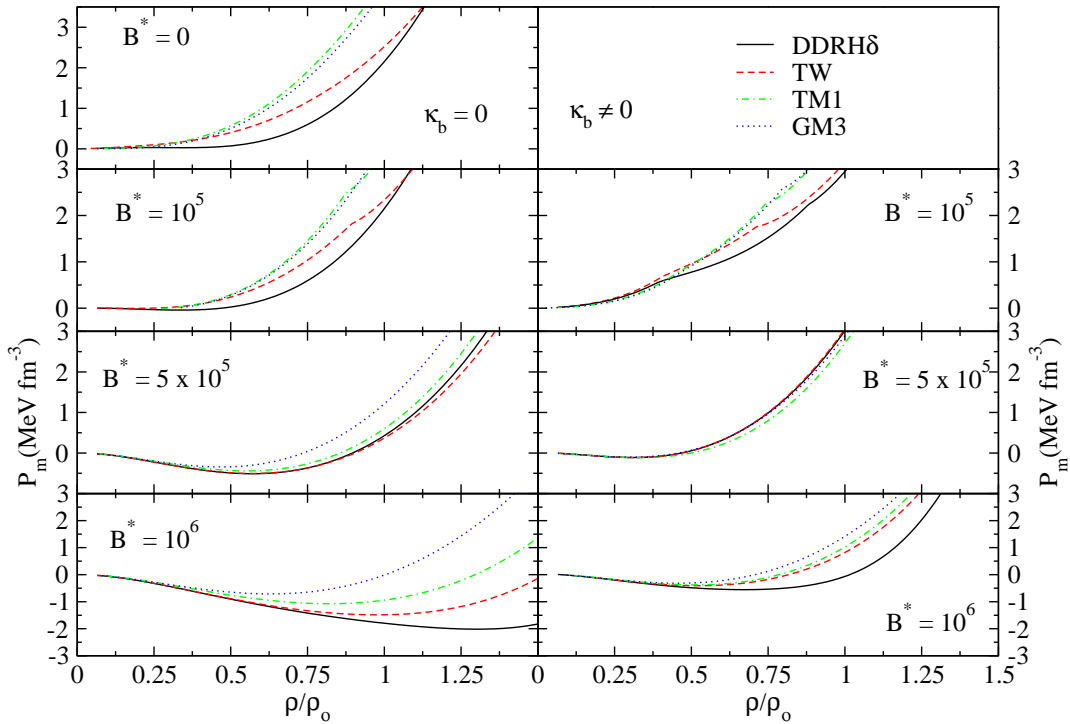}
\caption{(Color online)  EoS at low densities for stellar matter without (left) and with
(right) the nucleon anomalous magnetic moment. Several magnetic fields are
considered.}
\label{loweos}
\end{figure}

In Fig.~\ref{loweos} only the region $\rho<1.5 \rho_0$ is plotted. The magnetic field increases  the binding at low densities and a negative pressure may still occur beyond 2$\rho_0$. The most bound matter occurs for DDRH$\delta$ and this is due to the low symmetry energy this model has at intermediate densities. This means that the crust extends itself to higher densities and a higher value of the density will characterise  the inner edge of the crust. The inclusion of the AMM reduces this effect but binding still occurs and there are also regions of negative compressibility and/or pressure defining an unstable region. 
\begin{figure}[htb]
\centering
\includegraphics[width=0.75\linewidth,angle=-90]{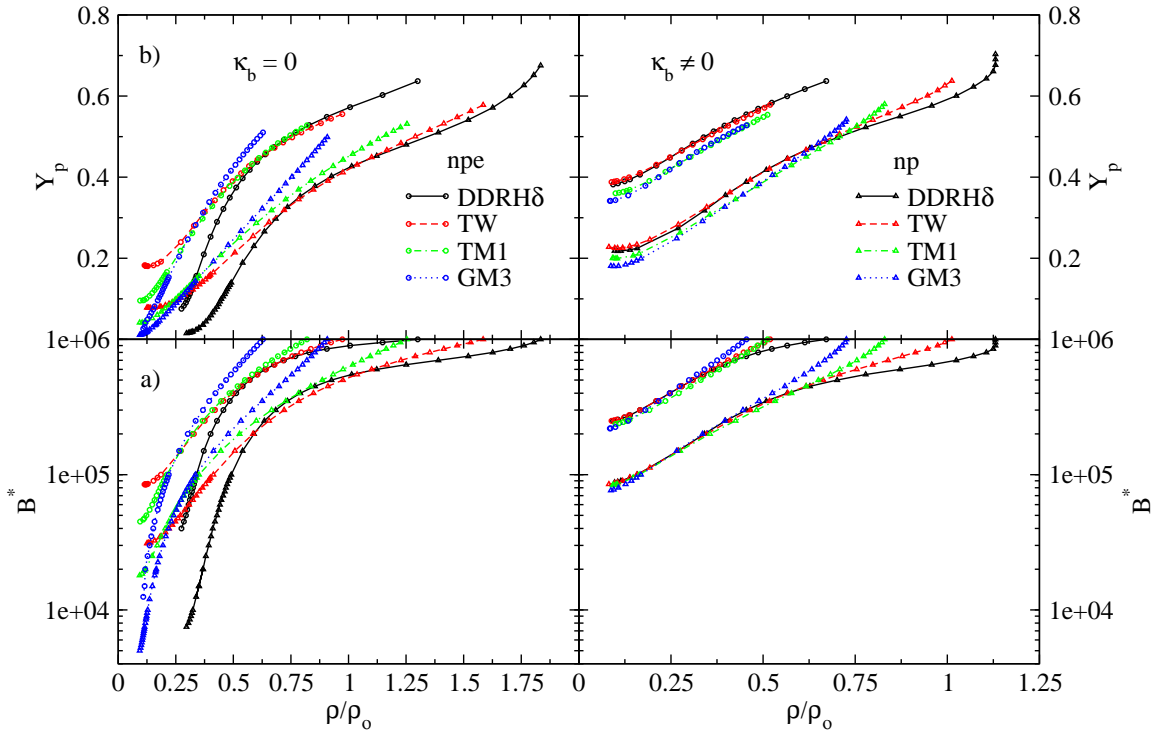}
\caption{(Color online)  Low density mechanical instability. Bottom:  The upper limit of the mechanical instability for different magnetic fields as a function of density; top: the corresponding  proton fraction; left: no AMM; right: including AMM.}
\label{inst}
\end{figure}
It is clearly seen that the models behave differently in these range of densities when the B field increases. GM3 becomes harder than TM1 for $B^*>5\times 10^5$. Also the relative behavior of TW and DDRH$\delta$ change.

In order to better understand the low density behavior we have determined the upper density limit of the mechanical instability, the density at which the incompressibility becomes zero, as a function of the magnetic field, Fig.~\ref{inst}a),  and calculated the associated proton fraction Fig.~\ref{inst}b). For $\beta$-equilibrium nuclear  matter with no magnetic field there are no  mechanical instabilities even if we only consider neutron-proton (np) matter and do not add the electron contribution, for which the incompressibility is always positive. In Table~\ref{table2} we give the densities for which the EoS for $\beta$-equilibrium  matter at a zero magnetic field  crosses the thermodynamical spinodal for $np$  matter.
As shown in~\cite{pasta08} these numbers give an order of magnitude of the upper limit of the transition
density: neither the Coulomb force nor the finite range of the nuclear force
is taken into account. They are slightly larger than the corresponding values obtained from the crossing of the dynamical spinodal with the EOS for neutron-proton-electron (npe) matter, which define a lower limit, and of the same order of magnitude of the results obtained from the transition of a pasta phase to a homogeneous phase~\cite{pasta08}.

\begin{table}[htb]
\caption{Predicted density and pressure at the inner edge
of the crust of a compact star at zero temperature,
as defined by the crossing between the thermodynamical instability region of np matter
and the $\beta$-equilibrium condition for homogeneous, neutrino-free stellar matter.}
\begin{tabular}{ccc}
\hline
\hline
 & $\rho_{b}(\hbox{fm}^{-3})$ & $P_{m}(\hbox{MeV}\hbox{fm}^{-3})$  \\
\hline
TM1                & 0.069509 & 0.50288   \\
\hline
GM3               & 0.068762  & 0.35644  \\
\hline
TW                  & 0.084955 & 0.52246 \\
\hline
DDRH$\delta$ & 0.085038 & 0.12855 \\
\hline\hline
\end{tabular}
\label{table2}
\end{table} 

From Fig.~\ref{inst} we see that the magnetic field, if strong enough, may change this picture, more strongly if the AMM is not considered. In  this figure we show both the mechanical instability upper bound for np matter and also for  npe matter. Of course in the last case the instability region is smaller due to the high incompressibility of the electron gas and occurs at larger proton fraction when the symmetry term contribution is smaller. The existence of a mechanical instability region in the presence of the magnetic field  has two reasons: a) due to the existence of Landau levels the nucleonic pressure does not increase so fast with density and b) at large magnetic fields the proton fraction increases and the symmetry repulsive term in the energy density is not so strong. The model DDRH$\delta$ is the one showing the largest instability ranges and at
least, for the no AMM calculation, it also predicts the smaller proton fractions. However, when the  AMM contribution is introduced the proton fraction of  all models behave in a very similar way except for the larger fields. In fact we should perform a  calculation which includes the Coulomb interaction and surface energy, but according to studies done in \cite{mueller95,abmp06,umodes06}, the  spinodal which includes these effects would be larger than  the mechanical instability region we have calculated.  A complete study of the low density region, namely the spinodal surface that
limits the non-homogeneous phase, needs to be done.

\begin{figure}[htb]
\centering
\includegraphics[width=0.75\linewidth,angle=-90]{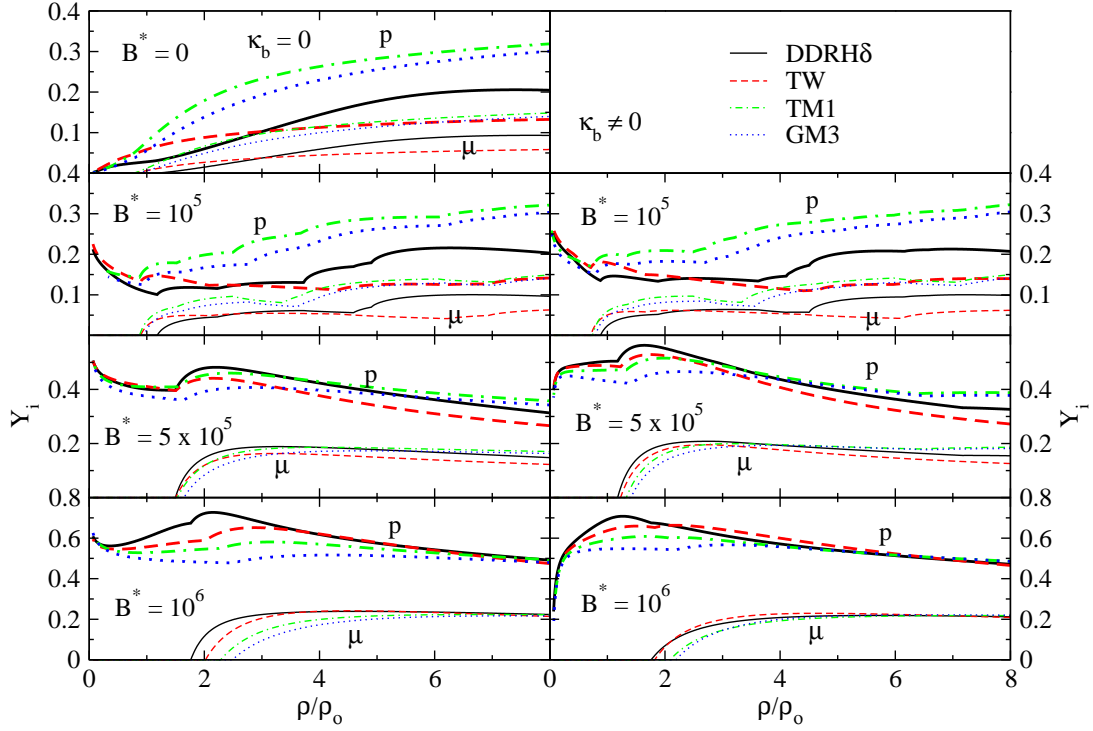}
\caption{(Color online)  Proton and muon fraction for stellar matter without (left) and with
(right) the nucleon anomalous magnetic moment. Several magnetic fields are
considered.}
\label{frac}
\end{figure}

\begin{figure}[htb]
\hspace*{-1.5cm}
\centering
\begin{minipage}[htb]{0.5\linewidth}
\includegraphics[width=0.875\linewidth,angle=-90]{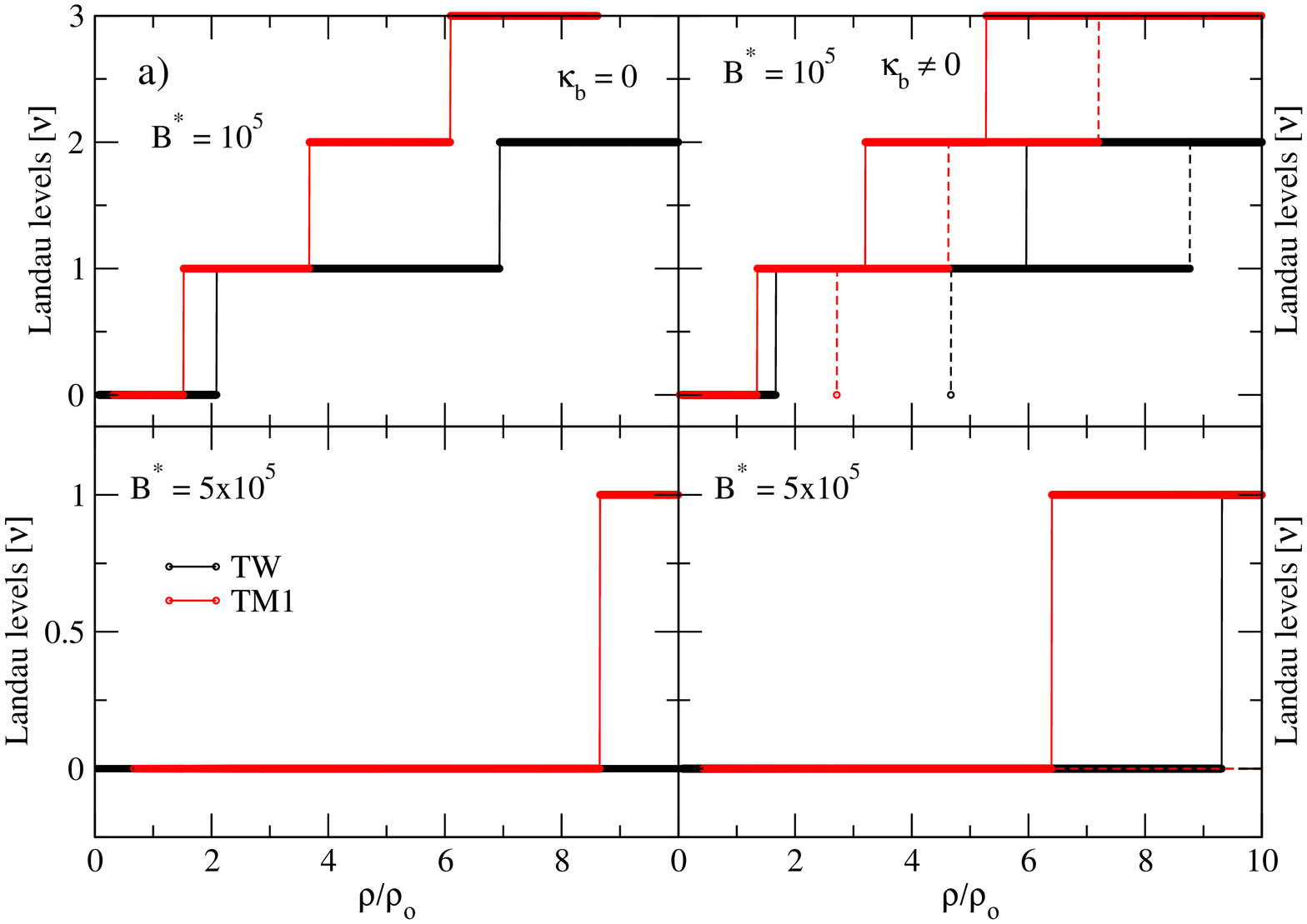}
\end{minipage}
\begin{minipage}[htb]{0.5\linewidth}
\hspace*{0.25cm}
\centering
\includegraphics[width=0.875\linewidth,angle=-90]{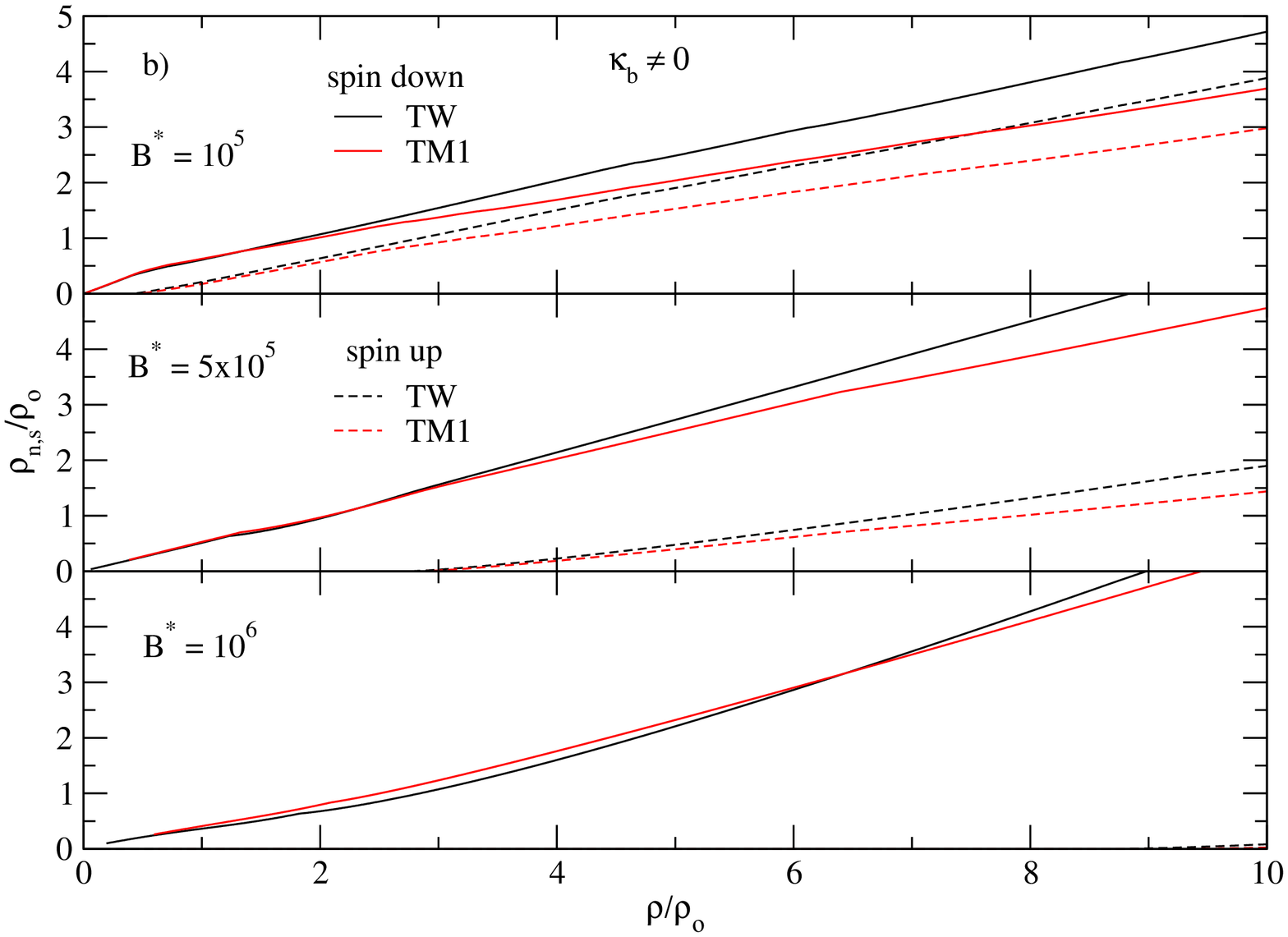}
\end{minipage}
\caption{(Color online) Density dependence of the proton and neutron polarization for TM1 and TW and different values of the magnetic field: a) Landau levels occupied by protons without (left) with (right) AMM. The dashed lines indicate the onset of spin down for each Landau level in the calculation with AMM; b) neutron polarization, full (dashed) lines correspond to spin down (up).}
\label{landau}
\end{figure}

The fraction of protons and muons  for the models and field intensities discussed are given in Fig.~\ref{frac} as a function of density. For comparison we also include the $B=0$ results. The effect of the magnetic field is  not very large for $B^*=10^5$, except at low densities, $\rho< 2 \rho_0$, when protons are totally polarised. In  Fig.~\ref{landau}a) we show the occupied Landau levels as a function of density and in Fig.~\ref{landau}b) the neutron polarization for the calculation including AMM. For $B^*=10^5$ ($B^*=5\times 10^5$) the second Landau level starts being occupied only for densities above $\sim 2 \rho_0$ ($\sim 8 \rho_0$). GM3 has a similar behaviour to TM1 and DDRH$\delta$ an intermediate behaviour between TM1 and TW. From Fig.~\ref{landau}b) we conclude that while at  $B^*=10^5$ neutrons are only slightly polarized at $B^*=10^6$ they are totally polarized for densities below $8\rho_0$. This total neutron polarization favours an increase of the proton fraction when AMM is included.

The fraction of protons within the different models is determined by the symmetry energy of the respective model for the density considered. This explains why for $B^* \le 10^5$, DDRH$\delta$ has the smallest fraction for $\rho<2.5 \rho_0$. For $\rho>2.5 \rho_0$ it is TW which has the smallest symmetry energy and the smallest proton fraction. For high fields, $B^*>10^6$ the fraction of protons is larger than the fraction of neutrons and again the symmetry energy defines the models with the largest fraction: DDRH$\delta$ for $\rho<3\rho_0$ and TW for densities larger than 3$\rho_0$. At $B^*= 5\times 10^5$ all proton fraction lie between 0.4 and 0.5 for densities below 4$\rho_0$. The relative fraction of protons is then determined by the effective mass: GM3 has the lower fraction due to its larger mass. On the other hand DDRH$\delta$ has the smallest proton mass a largest proton fraction. This effect is even larger for $B^*=10^6$. The inclusion of the AMM has important effects for $B^*>10^6$. For densities $\rho<2\rho_0$ the behaviour of the different models is still distinguishable but for larger densities the behaviour of all models is mostly determined by the magnetic field intensity. The magnetic field gives rise to an onset of muons at larger densities when the anomaly is not included because the magnetic fields favours a larger proton fraction and therefore a smaller electron chemical potential. With the introduction of the AMM,  the electron chemical potential increases slightly which explains the onset of muons at lower densities. 

\begin{figure}[htb]
\centering
\includegraphics[width=0.75\linewidth,angle=-90]{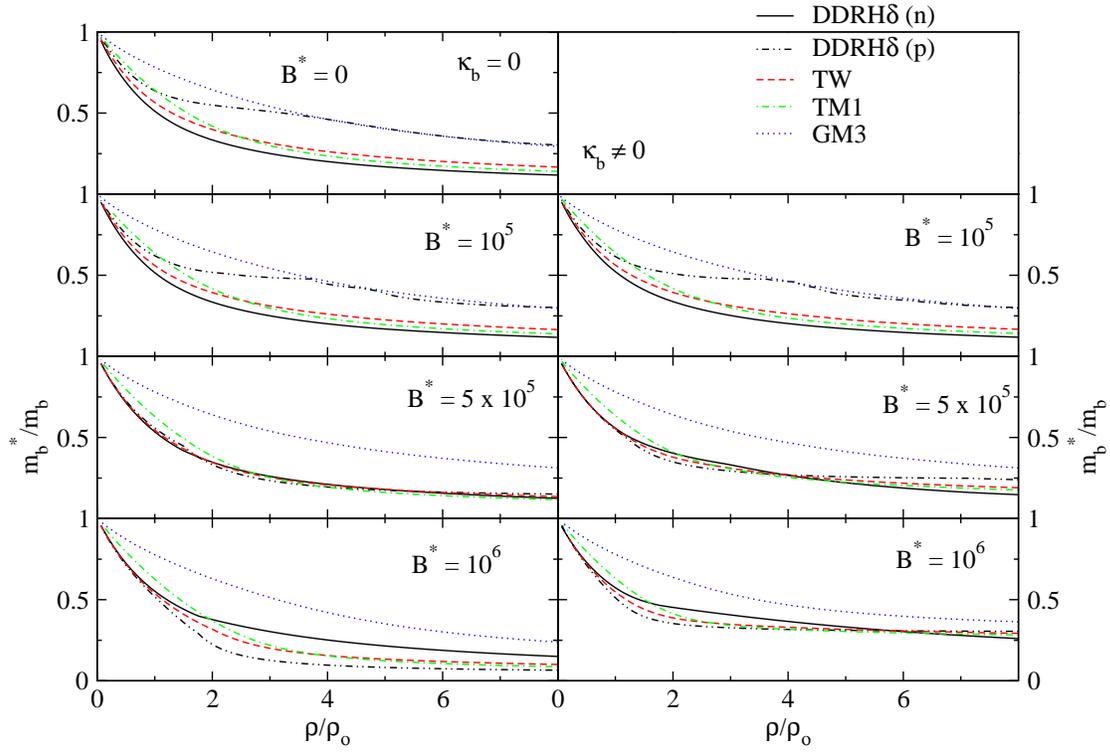}
\caption{(Color online)  Nucleon effective mass for stellar matter without (left) and with
(right) the nucleon anomalous magnetic moment.}
\label{meff}
\end{figure}

In Fig.~\ref{meff} the effective mass of the nucleons within the models under study are shown as a function of the density  for different intensities of the magnetic field, with and without the AMM. As discussed in~\cite{broderick}
when the AMM is not taken into account and the magnetic field becomes more intense, the effective mass
reduces faster with an increase of the  density. For $B^*=10^5$ the effect of the magnetic field is still small, but for larger values it becomes more important. For $B^*=10^5$ there is total polarisation only for $\rho< 2\rho_0$ while for $B^*=10^6$ there is total polarisation for the all range of densities considered. The GM3 model has the largest mass for all fields considered. At saturation and without magnetic field this model predicts an effective mass of 0.78 $M$ while for all the others the effective mass at saturation is 0.6 $M$ or smaller. The large values of the effective mass within GM3 justify the smaller proton fractions for the most intense magnetic fields. On the order hand DDRH$\delta$ shows the fastest reduction of the effective mass with density. For DDRH$\delta$ we show both the proton and the neutron mass. In this model, the mass of the most abundant nucleon, neutron for $B^*=10^5,\, 0.5\times 10^6$ and proton for $B^*=10^6$, or larger,  behaves like the nucleon mass in all the other models (except GM3), having slightly smaller values. The mass of the less abundant nucleon is quite higher, similar to the nucleon mass in GM3 for the smaller magnetic fields. If the AMM is included the effective masses saturate quite fast at a non-zero value which corresponds to  $m_b-s\mu_{N}\kappa_{i}B,\,\, i=p,n$. Models with density dependent couplings saturate faster than the other two.

The behaviour of the nucleon masses is better understood from the behaviour of  the scalar field  with density.
\begin{figure}[htb]
\centering
\includegraphics[width=0.75\linewidth,angle=-90]{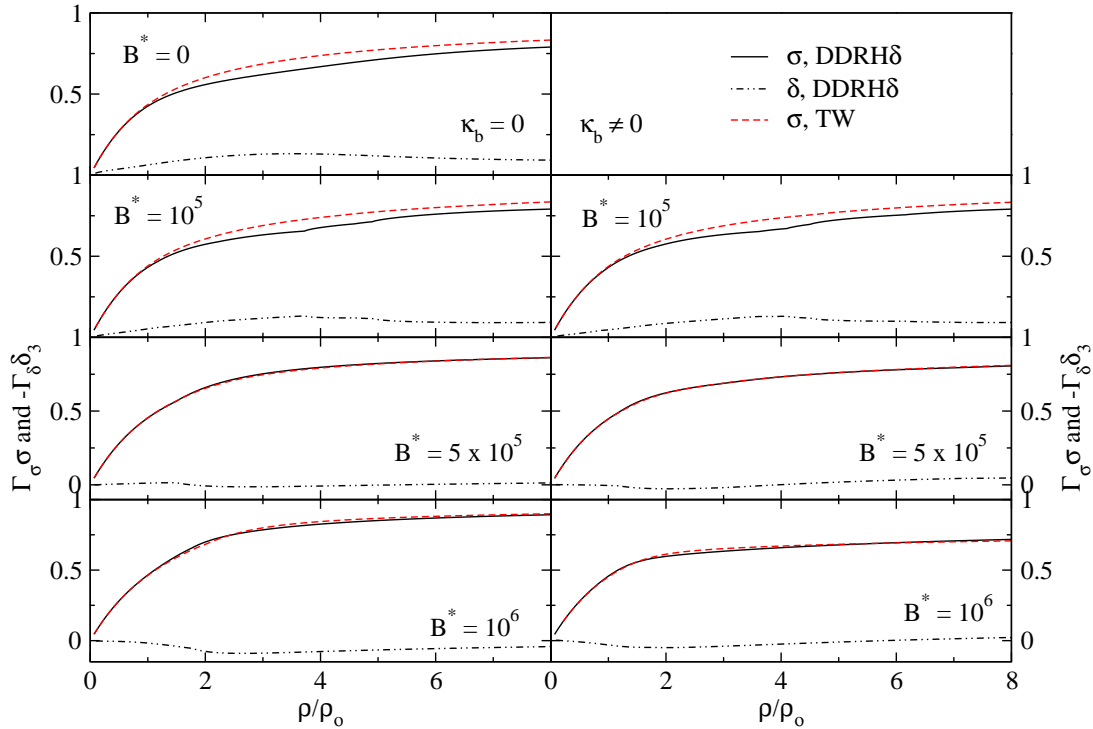}
\caption{(Color online)  The $\sigma$ and $\delta$-fields for stellar matter without (left) and with
(right) the nucleon anomalous magnetic moment.}
\label{fields}
\end{figure}
In Fig. \ref{fields} we plot the $\sigma$ and $\delta$ fields, more precisely $\Gamma_\sigma\sigma$ and $-\Gamma_\delta \delta_3$ as a function of the density for several magnetic field intensities. If the AMM is not included the $\delta$-field changes sign for the two most intensive fields considered. This reflects the existence of a  larger fraction of protons than neutrons. The inclusion of AMM reduces this effect and only for a restricted range of densities below $3-4\rho_0$. This is due to the larger effective mass of the protons. We also see that the $\sigma$ field increases faster with density for the
larger magnetic fields, giving rise to a faster saturation of the effective mass. If the AMM is taken into account the saturation of  $\sigma$ field does not occur at the baryonic mass value but at the value of the renormalised baryonic
mass by the AMM, see Eq. (\ref{barm}).

\section{Conclusions and outlooks}
In the present paper we have compared the EoS for stellar matter made out of
protons, neutrons, electrons and muons in the presence of very strong magnetic
fields. In particular, we have studied the EoS obtained within DDRH models with
and without the scalar-isovector meson $\delta$ and compared with other models
previously studied: GM3 and TM1.

It has been shown that, although the overall behaviour of all the models is similar, at low densities, $\rho<3\rho_0$,  the models show the largest differences. In particular, it has been shown that the low density instability region increases a lot as the magnetic field increases when the AMM is not taken into account. Although  the AMM reduces this behaviour there remains a region of instability not present for the magnetic field free matter. The larger range of instability is partially due to the larger proton fractions and partially due to the appearance of Landau levels.

Moreover DDRH$\delta$ model  shows a low density instability, which will give rise to a liquid-gas phase transition,  even for $B=0$. This could be due to the low symmetry energy it has at low and intermediate densities. It is also the  DDRH$\delta$ model that shows the largest changes of behaviour  with the magnetic field and the density,  both due to the presence of the $\delta$-meson and the density dependence of the coupling parameters. This is particularly clear with the EoS and the proton and muon fractions. In fact, the  DDRH models are the ones that predict smaller (higher) proton fractions at intermediate densities for $B^* \le 10^5 $ ($B^* \ge 10^6$), reflecting the behaviour of the symmetry energy,  the proton effective mass with density and the difference between proton and neutron masses.  Although the EoS of the different models do not differ so much, properties of the star sensitive to the proton fraction will distinguish the different models. These may be the neutrino interaction with hadronic matter, it is larger for larger neutron fractions, or the pairing properties of stellar matter which affect neutrino emissivities and specific heat. The stars more sensitive to the differences between the models are the low mass ones with M$\sim 1.0  M_\odot$. In particular, we have shown that the proton-neutron mass spitting present in DDR$\delta$ parametrization has noticeable effects on the proton fractions predicted by this model, changing at intermediate densities from the smallest ones to the largest ones.

In the present study we have not considered strangeness. At large densities
about two times the saturation density we may expect the onset of hyperons
\cite{glen00}, or kaon condensation \cite{glen99}. The effect of strong
magnetic fields on the onset of strangeness has been discussed in
\cite{broderick02,dey} and it was shown that both the hyperon  or the
kaon condensate onset occurs at larger densities, $>5\rho_0$, in the presence of strong
fields. So we may consider in the present discussion that results will not be
affected by the strangeness degree of freedom below 5 $\rho_0$. It remains,
however to be checked if the density dependence of the baryon-meson couplings
will have an effect on the hyperon onset.

The study of the effect of the magnetic field on the low density instabilities 
is of particular interest: the way the clusterization occurs and the extension
of the crust will affect the cooling and conduction properties of the star.
It was shown in the present work that the properties of the crust under strong
magnetic fields are sensitive to the EoS used. A detailed study of the effect
of strong magnetic fields on the  low density instabilities of nuclear
matter is  being carried on.

\begin{acknowledgments}

This work was partially supported by FEDER and FCT (Portugal)
under the grant SFRH/BPD/14831/2003, and
 projects POCI/FP/81923/2007 and  PDCT/FP/64707/2006. 
A. R. specially acknowledges many useful and elucidating discussions with A. M. Santos.

\end{acknowledgments}

\end{document}